\documentclass[a4paper,11pt]{article}
\usepackage{pos}

\title{Exotic heavy hadrons with a three-body nature}

\author*[a]{A. Mart\'inez Torres}
\author[a]{Brenda B. Malabarba}
\author[b]{Xiu-Lei Ren}
\author[c]{K. P. Khemchandani}
\author[d,e]{L. S. Geng}

\affiliation[a]{Universidade de Sao Paulo, Instituto de Fisica, C.P. 05389-970, Sao Paulo, Brazil.}
\affiliation[b]{Institut f\"ur Kernphysik $\&$ Cluster of Excellence PRISMA${}^{+}$, Johannes Gutenberg-Universität, D-55128 Mainz, Germany.}
\affiliation[c]{Universidade Federal de Sao Paulo, C.P. 01302-907, Sao Paulo, Brazil.}
\affiliation[d]{School of Physics, Beihang University, Beijing, 102206, China.}
\affiliation[e]{Beijing Key Laboratory of Advanced Nuclear Materials and Physics, Beihang University, Beijing, 102206, China}

\emailAdd{amartine@if.usp.br}

\abstract{In this talk we present a summary of our latest results on the investigation of three-body systems with explicit/hidden charm and with explicit/hidden strangeness. To be more concrete, in case of explicit strangeness quantum number, we pay attention on the $K D D $ and $K D\bar D^*$ systems, where, in the former, a charm $+2$, isospin $1/2$ and strangeness $+1$ state is obtained with a mass around 4140 MeV, while in the latter, a $K^*$ state, with hidden charm, and a mass close to 4307 MeV is found. In case of hidden strangeness, the $DK\bar K$ system is studied, while in the hidden charm and no strangeness quantum numbers sector, the $ND\bar D^*$ system is investigated. In the former a meson $D$ with mass around 2900 MeV is found to be generated, while in the latter several $N^*$ and $\Delta^*$ states with hidden charm are obtained with masses about $4600$ MeV. All these states constitute predictions of our model and future experimental confirmation of them would be relevant to elucidate the properties of the strong interaction in the presence of heavy quarks.}

\FullConference{%
  XV International Workshop on Hadron Physics (XV Hadron Physics)
  13 -17 September 2021 \\
  Online, hosted by Instituto Tecnológico de Aeronáutica, São José dos Campos, Brazil\\
  }

\begin{document}
\maketitle

\section{Introduction}
Within the present theory of Quantum Chromodynamics, nature allows the existence of exotic mesons and baryons, i.e., those whose quantum numbers can be reproduced with more than a $q\bar q$ seed for mesons and $qqq$ for baryons. Indeed, the formation of such hadrons has been confirmed experimentally, with some recent findings being those of the LHCb and BESIII collaborations related to the formation of multi-quark states with hidden charm, with and without strangeness~\cite{LHCb:2019kea,LHCb:2020bwg,BESIII:2020qkh}. Inside the category of exotic hadrons, particularly interesting are those with several units of charm and bottom quantum numbers. Such research would definitely be crucial in shedding light on the working of the strong interaction in the presence of heavy quarks. However, it is still not clear whether the properties of the exotic mesons and baryons found can be understood in terms of compact tetraquarks or as molecules of hadrons. And lot of theoretical and experimental efforts are being presently put to clarify this issue.

In this talk we summarize the results we found for several three-body systems involving heavy mesons, producing explicit and hidden charm as well as explicit and hidden strangeness.~In particular, we report our findings for the $KDD$, $KD\bar D^*$, $DK\bar K$ and $ND\bar D^*$ systems.

\section{Formalism and Results}
The solution of the Faddeev equations~\cite{Faddeev:1960su} to study three-body systems turns to be a quite useful method to investigate the formation of exotic mesons and baryons with a molecular nature. Considering as kernel for the Faddeev equations the two-body $t$-matrices obtained from the resolution of the Bethe-Salpeter equation in its on-shell factorization form~\cite{Oller:1997ti}, several three-body systems have been studied in the last years and generation of mesons and baryons with a molecular nature has been claimed~\cite{MartinezTorres:2007sr,Khemchandani:2008rk,MartinezTorres:2010zv}.

The advantage of this procedure is that not only the Bethe-Salpeter equation
\begin{align}
t=v+vgt,\label{BS}
\end{align}
where $v$ is the kernel and $g$ is a loop function of two hadrons, becomes an algebraic equation. The Faddeev equations
\begin{align}
T^1&=t^1+t^1 G[T^2+T^3],\nonumber\\
T^2&=t^2+t^2 G[T^1+T^2],\nonumber\\
T^3&=t^3+t^3 G[T^1+T^2],\label{Fa}\
\end{align}
where $t^i$, $i=1,2,3$ is the two-body $t$-matrix describing the interaction of particles $(jk)$, $j\neq k\neq i$ and $G$ is a three-body loop function, become algebraic equations too under certain approximations~\cite{MartinezTorres:2007sr,Khemchandani:2008rk,MartinezTorres:2010zv}. In this way, once the kernel $v$ in Eq.~(\ref{BS}) is obtained, the calculation of the three-body $T$-matrix for the system, $T=T^1+T^2+T^3$, gets tremendously simplified, even if a large number of coupled channels need to be considered.

In this approach, the kernel $v$ in Eq.~(\ref{BS}) is determined by using effective Lagrangians based on the relevant symmetries for the problem under investigation, like the chiral and heavy quark symmetries of the strong interaction~\cite{Gasser:1984gg,Georgi:1990um}. The solution of Eq.~(\ref{BS}) within coupled channels reveals the generation of two-hadron molecular states. In particular,  the $KD$, $\eta D_s$ interactions form the state $D^*_{s0}(2317)$~\cite{Kolomeitsev:2003ac,Guo:2006fu,Gamermann:2006nm,Torres:2014vna}. The $D\bar D^*$ system in isospin 0 generates $X(3872)$ while in isospin 1 the $Z_c(3900)$ is formed~\cite{Close:2003sg,Wong:2003xk,Swanson:2003tb,Gamermann:2007fi,Aceti:2014uea}. The $K\bar K$ interaction give rise to $f_0(980)$~\cite{Oller:1997ti}, while the $ND/ND^*$ system originates $\Lambda_c(2595)$~\cite{Garcia-Recio:2008rjt,Liang:2014kra,Nieves:2019nol}.

Considering these ingredients, the solution of Eq.~(\ref{Fa}) for the $DDK$, $DD\pi$ and $DD_s\eta$ coupled channel system shows the generation of a state with isospin 1/2, charm +2, strangeness +1 and mass around 4140 MeV~\cite{MartinezTorres:2018zbl}. Interestingly, this state arises when $D^*_{s0}(2317)$ is formed in one of the $DK$ subsystems.  As a consequence of this, the three-body state obtained can decay to two-body channels like $D_s D$, $D_s D^*$ and $D D^*_s$ through triangular loops, producing a small width. These decay mechanisms were investigated in Ref.~\cite{Huang:2019qmw}, finding a total width for the state of 2-3 MeV. Preliminary search for a state with charm +2 and mass around 4100 MeV has been recently conducted by the Belle collaboration~\cite{Belle:2020xca} and no clear signal was found, however, more precise data are necessary to get better information about the existence of such a state.

In case of the $KD\bar D^*$ system, a $K^*$ meson with mass around 4307 MeV is obtained from the resolution of Eq.~(\ref{Fa}). In this case, the three-body state is generated when the $KD$ system in isospin 0 forms $D^*_{s0}(2317)$ and the $D\bar D^*$ system produces $X(3872)$ in isospin 0 and $Z_c(3900)$ in isospin 1~\cite{Ren:2018pcd}. Since in our formalism $Z_c(3900)$ can be considered as a state generated from the $D\bar D^*$ and $J/\psi\pi$ coupled channels, it can decay to $J/\psi\pi$~\cite{Tanabashi:2018oca}, producing a width for $Z_c(3900)$ of around 30 MeV. When implementing this width into our three-body calculation, the $K^*(4307)$ state obtains a width of around 18 MeV. In view of the nature found for this three-body state, $K^*(4307)$ can decay to two-body channels too, like $J/\psi K^*(892)$, $\bar D D_s$, $\bar D D^*_s$ and $\bar D^* D^*_s$, through triangular loops. These decay widths were determined in Ref.~\cite{Ren:2019umd}, finding $\sim7$ MeV for $J/\psi K^*(892)$, $\sim0.5$ for $\bar D D^*_s$, $\bar D^* D^*_s$ and $\sim1$ MeV for $\bar D D_s$. The fact of having an important $KZ_c(3900)$ molecular component in the wave function of $K^*(4307)$ indicates that the state can also easily decay to a channel like $J/\psi\pi K$.  In this way, the reconstruction on the $J/\psi\pi K$ invariant mass in processes involving these particles in their final states can be a promising way of observing this $K^*(4307)$~\cite{Ren:2019rts}.

For the $DK\bar K$ system, by solving Eq.~(\ref{Fa}), a $D$ meson with mass around 2900 MeV and width of 55 MeV is found~\cite{MartinezTorres:2012jr} when the $K\bar K$ system generates $f_0(980)$. A state with such an internal structure can decay to two-body channels like $D^*\pi$, $D^*\bar K$, $D^*_{s0}\bar K$, out of which, the largest decay width comes from $D(2900)\to D^*_{s0}(2317)\bar K$~\cite{Malabarba:2021gyq}. Structures around 3000 MeV have been observed by the LHCb collaboration in the $D^*\pi$ and $D\pi$ invariant masses~\cite{LHCb:2013jjb} and could correspond to the $D(2900)$ predicted~\cite{Malabarba:2021gyq}.

In the hidden charm and null strangeness sector, the study of the $ND\bar D^*$ system reveals formation of several narrow $N^*$ and $\Delta ^*$ states with masses around $4400-4600$ MeV and a positive parity. In this case, the states are obained when the $D\bar D^*$ system generates $X(3872)$ and $Z_c(3900)$ and the $ND/ND^*$ system cluster as $\Lambda_c(2595)$~\cite{Malabarba:2021taj}. The investigation of two-body decay channels like $N J/\psi$ and $\bar D^{(*)}\Sigma_c$, which is a consequence of the nature found for these three-body states, can be useful in finding signals for such exotic hadrons.

\section{Conclussions}
In this talk we have presented our latest results for the generation of exotic mesons and baryons as a consequence of the three-body dynamics involved in several systems constituted by heavy mesons, like $KDD$, $KD\bar D^*$, $DK\bar K$ and $ND\bar D^*$. In particular, in the $KDD$ system we find a state around 4140 MeV with charm $+2$ and strangeness $+1$. The study of the $KD\bar D^*$ and $ND\bar D^*$ systems reveals the formation of Kaon and Nucleon/Delta resonances with hidden charm, respectively, with masses in the region of 4000-4500 MeV. In the hidden strangeness sector, a $D$ meson with mass about 2900 MeV is obtained as a consequence of the dynamics involved in the $DK\bar K$ system.

\acknowledgments
This work is supported by the Funda\c c\~ao de Amparo \`a Pesquisa do Estado de S\~ao Paulo (FAPESP), processos n${}^\circ$ 2019/17149-3, 2019/16924-3 and 2020/00676-8, and by the Conselho Nacional de Desenvolvimento Cient\'ifico e Tecnol\'ogico (CNPq), grant  n${}^\circ$ 305526/2019-7 and 303945/2019-2.


\begin{thebibliography}{99}
\bibitem{LHCb:2019kea}
R.~Aaij \textit{et al.} [LHCb],
Phys. Rev. Lett. \textbf{122}, no.22, 222001 (2019).

\bibitem{LHCb:2020bwg}
R.~Aaij \textit{et al.} [LHCb],
Sci. Bull. \textbf{65}, no.23, 1983-1993 (2020).

\bibitem{BESIII:2020qkh}
M.~Ablikim \textit{et al.} [BESIII],
Phys. Rev. Lett. \textbf{126}, no.10, 102001 (2021).
      
    
 \bibitem{Faddeev:1960su} 
  L.~D.~Faddeev,
  {\it Sov.\ Phys.\ JETP} {\bf 12}, 1014 (1961)
  [{\it Zh.\ Eksp.\ Teor.\ Fiz.}\  {\bf 39}, 1459 (1960)].
  
   \bibitem{Oller:1997ti} 
  J.~A.~Oller and E.~Oset,
  {\it Nucl.\ Phys.\ A} {\bf 620}, 438 (1997)
  Erratum: [{\it Nucl.\ Phys.\ A} {\bf 652}, 407 (1999)].
  
  \bibitem{MartinezTorres:2007sr} 
  A.~Martinez Torres, K.~P.~Khemchandani and E.~Oset,
  {\it Phys.\ Rev.\ C} {\bf 77}, 042203 (2008).
  
  \bibitem{Khemchandani:2008rk} 
  K.~P.~Khemchandani, A.~Martinez Torres and E.~Oset,
  {\it Eur.\ Phys.\ J.\ A} {\bf 37}, 233 (2008).
  
  \bibitem{MartinezTorres:2010zv} 
  A.~Martinez Torres and D.~Jido,
  {\it Phys.\ Rev.\ C} {\bf 82}, 038202 (2010).
    
     
  \bibitem{Gasser:1984gg} 
  J.~Gasser and H.~Leutwyler,
  {\it Nucl.\ Phys.\ B} {\bf 250}, 465 (1985).
  
  \bibitem{Georgi:1990um} 
  H.~Georgi,
  {\it Phys.\ Lett.\ B} {\bf 240}, 447 (1990).
 
   
  \bibitem{Kolomeitsev:2003ac} 
  E.~E.~Kolomeitsev and M.~F.~M.~Lutz,
  {\it Phys.\ Lett.\ B} {\bf 582}, 39 (2004).
  
  \bibitem{Guo:2006fu} 
  F.~K.~Guo {\it et. al.},
  {\it Phys.\ Lett.\ B} {\bf 641}, 278 (2006).
  
  \bibitem{Gamermann:2006nm} 
  D.~Gamermann, E.~Oset, D.~Strottman and M.~J.~Vicente Vacas,
  {\it Phys.\ Rev.\ D} {\bf 76}, 074016 (2007).
  
  \bibitem{Torres:2014vna} 
  A.~Martinez Torres, E.~Oset, S.~Prelovsek and A.~Ramos,
  {\it JHEP} {\bf 1505}, 153 (2015).
  
   \bibitem{Close:2003sg} 
  F.~E.~Close and P.~R.~Page,
  {\it Phys.\ Lett.\ B} {\bf 578}, 119 (2004).
  
  \bibitem{Wong:2003xk} 
  C.~Y.~Wong,
  {\it Phys.\ Rev.\ C} {\bf 69}, 055202 (2004).
  
  \bibitem{Swanson:2003tb} 
  E.~S.~Swanson,
  {\it Phys.\ Lett.\ B} {\bf 588}, 189 (2004).
  
  \bibitem{Gamermann:2007fi} 
  D.~Gamermann and E.~Oset,
  {\it Eur.\ Phys.\ J.\ A} {\bf 33}, 119 (2007).
  
    
  \bibitem{Aceti:2014uea} 
  F.~Aceti {\it et. al}.,
  {\it Phys.\ Rev.\ D} {\bf 90}, no. 1, 016003 (2014).
  
  
  
  \bibitem{Garcia-Recio:2008rjt}
C.~Garcia-Recio, V.~K.~Magas, T.~Mizutani, J.~Nieves, A.~Ramos, L.~L.~Salcedo and L.~Tolos,
Phys. Rev. D \textbf{79}, 054004 (2009).

  \bibitem{Liang:2014kra}
W.~H.~Liang, T.~Uchino, C.~W.~Xiao and E.~Oset,
Eur. Phys. J. A \textbf{51}, no.2, 16 (2015).

  \bibitem{Nieves:2019nol}
J.~Nieves and R.~Pavao,
Phys. Rev. D \textbf{101}, no.1, 014018 (2020).

  
  \bibitem{MartinezTorres:2018zbl} 
  A.~Martinez Torres, K.~P.~Khemchandani and L.~S.~Geng,
  Phys.\ Rev.\ D {\bf 99}, no. 7, 076017 (2019).
    
  \bibitem{Huang:2019qmw} 
  Y.~Huang, M.~Z.~Liu, Y.~W.~Pan, L.~S.~Geng, A.~Martínez Torres and K.~P.~Khemchandani,
  Phys. Rev. D \textbf{101}, no.1, 014022 (2020).
  
  \bibitem{Belle:2020xca}
Y.~Li \textit{et al.} [Belle],
Phys. Rev. D \textbf{102}, no.11, 112001 (2020).


   \bibitem{Ren:2018pcd} 
  X.~L.~Ren, B.~B.~Malabarba, L.~S.~Geng, K.~P.~Khemchandani and A.~Martinez Torres,
  Phys.\ Lett.\ B {\bf 785}, 112 (2018).

  \bibitem{Tanabashi:2018oca} 
  M.~Tanabashi {\it et al.},
  {\it Phys.\ Rev.\ D} {\bf 98}, no. 3, 030001 (2018).
  
  \bibitem{Ren:2019umd} 
  X.~L.~Ren, B.~B.~Malabarba, K.~P.~Khemchandani and A.~Martinez Torres,
  JHEP {\bf 1905}, 103 (2019).
  
  \bibitem{Ren:2019rts}
X.~L.~Ren, K.~P.~Khemchandani and A.~Martinez Torres,
Phys. Rev. D \textbf{102}, no.1, 016005 (2020).
  
    

  \bibitem{Xie:2010ig} 
  J.~J.~Xie, A.~Martinez Torres and E.~Oset,
  {\it Phys.\ Rev.\ C} {\bf 83}, 065207 (2011).
  
  \bibitem{Xie:2011uw} 
  J.~J.~Xie {\it et. al.},
  {\it Phys.\ Rev.\ C} {\bf 83}, 055204 (2011).

  \bibitem{MartinezTorres:2010ax} 
  A.~Martinez Torres, E.~J.~Garzon, E.~Oset and L.~R.~Dai,
  {\it Phys.\ Rev.\ D} {\bf 83}, 116002 (2011).
  
  \bibitem{MartinezTorres:2012jr}
A.~Martinez Torres, K.~P.~Khemchandani, M.~Nielsen and F.~S.~Navarra,
Phys. Rev. D \textbf{87}, no.3, 034025 (2013).
  
  \bibitem{Malabarba:2021gyq}
B.~B.~Malabarba, K.~P.~Khemchandani and A.~M.~Torres,
Phys. Rev. D \textbf{104}, no.11, 116002 (2021).
  
  \bibitem{LHCb:2013jjb}
R.~Aaij \textit{et al.} [LHCb],
JHEP \textbf{09}, 145 (2013).

\bibitem{Malabarba:2021taj}
B.~B.~Malabarba, K.~P.~Khemchandani and A.~M.~Torres,
Eur. Phys. J. A \textbf{58}, no.2, 33 (2022).

\end{thebibliography}
\end{document}